\documentclass[a4paper,12pt]{article}
\usepackage{graphicx,amssymb}
\renewcommand{\figurename}{Fig.}

\title{Giant neutron halo in nuclei beyond beta-stability line}

\author{V. M. Kolomietz, S. V. Lukyanov, A. I. Sanzhur \\
{\it Institute for Nuclear Research, 03680 Kyiv, Ukraine}}

\date{\empty}

\begin{document}
\maketitle

\begin{abstract}
The radii of nucleon distribution and neutron skin in nuclei beyond the $\beta$-stability line are studied 
within the extended Thomas-Fermi approximation. We show that the growth of neutron skin in unstable nuclei 
does not obey the saturation condition because of the neutron coat. The neutron coat indicates the possibility 
of giant neutron halo which is growing with moving away from the beta stability line. We demonstrate the 
presence of strong shell oscillations in the charge radius $R_{C}$ and the relation of $R_{C}$ to the isospin 
shift of  neutron-proton chemical potentials $\Delta\lambda =\lambda_{n}-\lambda_{p}$ for nuclei beyond the 
beta-stability line at fixed value of mass number $A$.
\bigskip


\end{abstract}

\section{Introduction}

In the vicinity of the beta-stability line, the average changes in binding energy $E$ and nuclear radius $R$ 
with nucleon content obey the saturation properties. The volume part $E_{\mathrm{vol}}$ of binding energy and 
the nuclear volume itself are proportional to the particle number $A$ with $E_{\mathrm{vol}}=-b_{V}A$ and $R=r_{0}A^{1/3}$, where $b_{V}>0$ and $r_{0}$ and are constants. Both values of $b_{V}$ and $r_{0}$ depend, 
however, on the isotopic asymmetry parameter $X=(N-Z)/(N+Z)$. This dependence comes from the difference in 
saturation bulk density, $\rho_{0}\sim r_{0}^{-3}$, for nuclei with different values of $X$. The saturation 
density $\rho_{0}$ becomes smaller beyond the beta-stability line for neutron-rich nuclei where more neutrons 
are pushed off to form the "neutron coat". One can expect that the growth of neutron skin in neutron-rich nuclei
violates the saturation property $R\sim A^{1/3}$ for the nuclear radius providing an existence of neutron halo 
(giant neutron halo) effect \cite{meto02}. 

In this paper we study the deviation of neutron distribution from the saturation behavior in neutron-rich nuclei. 
We study the influence of spin-orbit and Coulomb forces on the neutron, $\sqrt{\left\langle r_{n}^{2}\right\rangle}$,
and proton, $\sqrt{\left\langle r_{p}^{2}\right\rangle }$, root mean square radii as well as the relation of the 
shift $\sqrt{\left\langle r_{n}^{2}\right\rangle}-\sqrt{\left\langle r_{p}^{2}\right\rangle}$ to the surface 
symmetry energy. We study the problems related to the nucleon redistribution within the surface region of the 
nucleus and, in particular, the neutron coat and the neutron excess for the nuclei beyond the beta stability 
line. 

We combine the extended Thomas-Fermi approximation (ETFA) and the direct variational method assuming that the 
proton and neutron distributions are sharp enough, i.e., that the corresponding densities $\rho_{p}(\mathbf{r})$
and $\rho_{n}(\mathbf{r})$ fall from their bulk values to zero within a thin surface region. In our consideration, 
the thin-skinned densities $\rho_{p}(\mathbf{r})$ and $\rho_{n}(\mathbf{r})$ are generated by the profile functions
which are eliminated by the requirement that the energy of the nucleus should be stationary with respect to 
variations of these profiles.

\section{Direct variational approach}
We will use the ETFA which is one of practical realization of general 
Hohenberg-Kohn theorem \cite{kosh64} on the unique functional relation between the ground state energy and 
the local density of particles for any fermion system. The total kinetic energy of the many-body fermion system 
is given by the semiclassical expression \cite{kirzh,book} as follows 
\begin{equation}
E_{\mathrm{kin}}\{\rho_{n},\rho_{p}\}
\equiv E_{\mathrm{kin}}\{\rho _{q},\mathbf{\nabla }\rho_{q}\}
=\int d\mathbf{r}\,\,\epsilon_{\mathrm{kin}}[\rho_{n}(\mathbf{r}),\rho_{p}(\mathbf{r})],  
\label{ekin}
\end{equation}
where $\epsilon_{\mathrm{kin}}[\rho_{n},\rho_{p}]
=\epsilon_{\mathrm{kin,}n}[\rho_{n}]+\epsilon_{\mathrm{kin,}p}[\rho_{p}]
$, and 
$$
\epsilon_{\mathrm{kin,}q}[\rho_{q},\mathbf{\nabla}\rho_{q}]
={\frac{\hbar^{2}}{2m}}\left[{\frac{3}{5}}\,(3\,\pi ^{2})^{2/3}\,\rho_{q}^{5/3}
+{\frac{1}{36}\frac{(\mathbf{\nabla}\rho_{q})^{2}}{\rho_{q}}}+{\frac{1}{3}}\,\nabla^{2}\rho_{q}\right].
$$
Here $\rho_{q}$ is the nucleon density with $q=n$ for neutron and $q=p$ for proton. 

We will follow the concept of effective nucleon-nucleon interaction using the Skyrme-type force. The total 
energy functional for charged nucleus is given by 
\begin{equation}
E_{\mathrm{tot}}\{\rho_{q},\mathbf{\nabla}\rho_{q}\}
=E_{\mathrm{kin}}\{\rho_{q},\mathbf{\nabla}\rho_{q}\}+E_{\mathrm{SK}}\{\rho_{q},\mathbf{\nabla}
\rho_{q}\}+E_{\mathrm{C}}\{\rho_{p}\},  
\label{etot}
\end{equation}
where $E_{\mathrm{SK}}\{\rho_{q},\mathbf{\nabla}\rho_{q}\}$ is the potential energy of $NN$-interaction
\begin{equation}
E_{\mathrm{SK}}\{\rho_{q},\mathbf{\nabla}\rho_{q}\}
=\int d\mathbf{r} \,\,\epsilon_{\mathrm{pot}}[\rho_{n}(\mathbf{r}),\rho_{p}(\mathbf{r})],
\label{epot}
\end{equation}                                          
$\epsilon_{\mathrm{pot}}[\rho_{n}(\mathbf{r}),\rho_{p}(\mathbf{r})]$ is the potential energy density and  
is the Coulomb energy. The potential energy (\ref{epot}) also includes the energy of spin-orbit interaction.
Considering the asymmetric nuclei with $X=(N-Z)/A\ll 1$, we will introduce the isotopic particle densities, 
namely the total density $\rho_{+}=\rho_{n}+\rho_{p}$ and the neutron excess density $\rho_{-}=\rho_{n}-\rho_{p}$ 
with $\rho_{-}\ll\rho_{+}$. We apply the direct variational method \cite{kosa08} and assume the density 
profile functions $\rho_{+}(r)$ and $\rho_{-}(r)$ to be a power of the Fermi function as 
\begin{equation}
\rho_{+}(r)=\rho_{0}\ f(r), \quad 
\rho_{-}(r)=\rho_{1}\ f(r)-\frac{1}{2}\rho_{0}\ \frac{df(r)}{dr}\Delta .  
\label{prof1}
\end{equation}
Here, $f(r)=\left[1+\exp\left[(r-R)/a\right]\right]^{-\eta}$, the values $\rho_{0}$ and $\rho_{1}$ are related 
to the bulk density, $R$ is the nuclear radius, $a$ is the diffuseness parameter and $\Delta$ is the parameter 
of neutron skin (see below). The profile functions $\rho_{+}(r)$ and $\rho_{-}(r)$ have to obey the condition 
of the neutron and proton number conservation. For the ground state of nucleus, the unknown parameters $\rho_{0,1}$,
$R$, $a$, $\Delta$, $\eta$ and the total energy $E_{\mathrm{tot}}$ itself can be derived from the variational principle 
\begin{equation}
\delta (E-\lambda_{n}N-\lambda_{p}Z)=0,  
\label{var1}
\end{equation}
where the variation with respect to all possible small changes of $\rho_{0,1}$, $R$, $a$, $\Delta$ and $\eta$ 
is assumed. The Lagrange multipliers $\lambda_{n}$ and $\lambda_{p}$ are the chemical potentials for neutrons 
and protons respectively, and both of them are fixed by the condition of particle number conservation. 

As mentioned above, the parameter $\Delta$ in profile functions of Eq. (\ref{prof1}) is related to the neutron 
skin. It can be easily seen from the derivation of the rms radii of the neutron and proton density distributions
\begin{equation}
\sqrt{\left\langle r_{q}^{2}\right\rangle}
=\sqrt{\int {d\mathbf{r}\,r^{2}\,\rho _{q}(r)}\left/ \int {d\mathbf{r}\,\,\rho_{q}(r)}\right.}.
\label{rms}
\end{equation}
Using Eqs. (\ref{rms}) and (\ref{prof1}) one obtains the size of the neutron skin as
\begin{equation}
\sqrt{\left\langle r_{n}^{2}\right\rangle }-\sqrt{\left\langle r_{p}^{2}\right\rangle}
\approx \sqrt{\frac{3}{5}}\frac{\Delta}{1-X^{2}}\left(1+\frac{\Delta}{2R}\frac{X\ }{1-X^{2}}\right) 
+O\left( \left( \frac{a}{R}\right)^{3}\right).  
\label{rms1}
\end{equation}
Note that the evaluation of the variational conditions leads to an additional dependence of the variational 
parameters $\rho_{0,1}$, $R$, $a$, $\Delta$ and $\eta$ on the external parameters $A$ and $X$.
The value of $\Delta$ disappears in symmetric nuclei at $X=0$ and depends slightly on the Skyrme force parametrization. In case of the SkM forces we have numerically calculated the dependence of
$\Delta$ on $X$ for $A=120$ and fitted it by the following formula
\begin{equation}
\Delta (X)\approx 0.90\ X+1.47\ X^{2}.  
\label{delta1}
\end{equation}

\begin{figure}
\begin{center}
\includegraphics*[scale=0.5,clip]{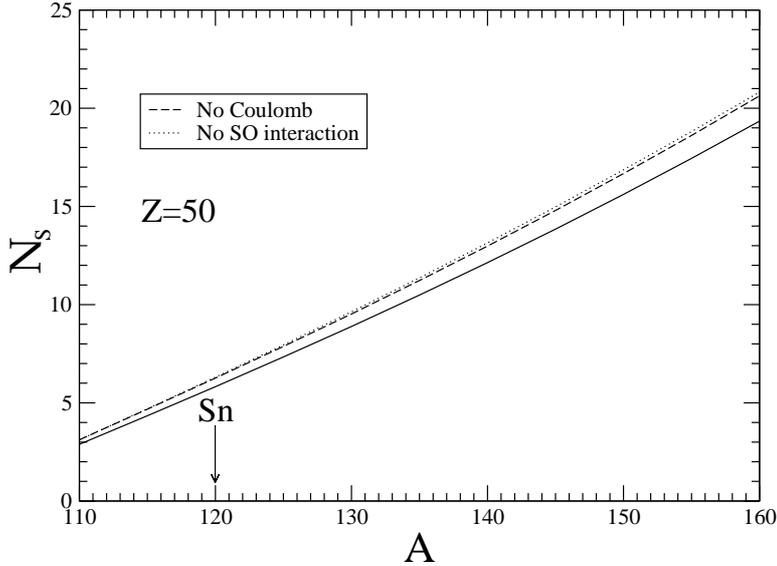}
\end{center}
\caption{$N_S$ for neutron-rich nuclei in the vicinity of Sn (Z=50) (solid line). The dotted and dashed lines
show the influence of the spin-orbit and the Coulomb forces on the neutron coat $N_S$. The arrow shows the 
position of Sn nucleus on the $\beta$-stability line.}
\label{fig1}
\end{figure}

The parameter $\Delta$ is also related to the number, $N_S$, of neutrons in surface region of the nucleus 
("neutron coat"). Substituting Eqs. (\ref{prof1}) into condition of the particle conservation and using the leptodermous expansion, we obtain for the neutron excess $N-Z$ the following expression 
\begin{equation}
N-Z\approx N_{V}+N_{S},  
\label{NZ1}
\end{equation}
where
$$
N_{V}
=\frac{4\pi}{3}R^{3}\left(1+3\kappa_{0}(\eta)\frac{a}{R}+6\kappa_{1}(\eta)\frac{a^{2}}{R^{2}}\right)\rho_{1},
\quad
N_{S}
=4\pi R^{2}\left(1+2\kappa_{0}(\eta)\frac{a}{R}+2\kappa_{1}(\eta)\frac{a^{2}}{R^{2}}\right)\frac{\rho _{0}}{2}\Delta
$$
and $\kappa_{i}(\eta)$ are the generalized Fermi integrals derived in Ref. \cite{kosa08}. The first term 
$N_{V}\sim R^{3}$ on the right hand side of Eq. \ref{NZ1}) is due to redistribution of the neutron excess within 
the nuclear volume while the second one $N_{S}\sim R^{2}$ is the number of neutrons within neutron coat. In 
\figurename\ \ref{fig1} we have plotted $N_S$ for neutron-rich nuclei in the vicinity of Sn nucleus (Z=50) 
(solid line). 
Two additional lines show the influence of the spin-orbit and the Coulomb forces on the neutron coat $N_S$.

In general, the change of the radius $R$ of nucleon distribution with the nucleon number $A$ is caused by two 
factors. There is a simple geometrical change $R\propto A^{1/3}$. An additional change can occur due to the
polarization effect (the bulk density distortion) with moving away from the beta-stability line. In particular, 
the size of neutron skin is sensitive to the symmetry and the Coulomb energies. We expand the total energy 
$E_{\mathrm{tot}}(\rho_{0},X)/A$ around the saturation density $\rho_{0,\mathrm{eq}}$ and the isotopic asymmetry
parameter $X^*$ on the beta-stability line as 
\begin{equation}
E_{\mathrm{tot}}(\rho_{0},X)/A
=E_{\mathrm{tot}}(\rho _{0,\mathrm{eq}},X^{\ast })/A
+\frac{K_A}{18\rho_{0,\mathrm{eq}}^{2}}(\rho_{0}-\rho_{0,\mathrm{eq}})^{2}
+\frac{P_{A}}{\rho_{0,\mathrm{eq}}^{2}}(X-X^{\ast})^{2}(\rho_{0}-\rho_{0,\mathrm{eq}}),  \label{e3}
\end{equation}
where $K_A$ is the incompressibility of finite nucleus
\begin{equation}
K_{A}=9\left. \rho_{0,\mathrm{eq}}^{2}\frac{\partial^{2}E_{\mathrm{tot}}(\rho_{0},X^{\ast })/A}
{\partial\rho_{0}^{2}}\right\vert_{A,\rho_{0}=\rho_{0,\mathrm{eq}}}  
\label{ka}
\end{equation}
and $P_A$ is the partial pressure related to the symmetry and the Coulomb energies
\begin{equation}
\left.P_{A}
=\rho_{0,\mathrm{eq}}^{2}\frac{\partial}{\partial\rho_{0}}\left[b_{\mathrm{V,sym}}(\rho_{0})
+b_{\mathrm{S,sym}}(\rho_{0})\ A^{-1/3}-\alpha_{C}(\rho_{0})A^{2/3}\right]
\right\vert_{\rho_{0}=\rho_{0,\mathrm{eq}}},
\label{pa}
\end{equation}
where $b_{\mathrm{V,sym}}(\rho_{0})$ and $b_{\mathrm{S,sym}}(\rho_{0})$ are the volume and the surface symmetry
coefficients and $\alpha_{C}(\rho_{0})=3e^2/20\ \left(4\pi\rho_0/3\right)^{1/3}$. As seen from Eq. (\ref{e3}), 
the deviation from the beta-stability line ($X\neq X^{\ast}$) implies the change in the bulk density $\rho_{0}$. 
The corresponding change is dependent on the incompressibility $K_A$ and the partial pressure $P_A$. For an 
arbitrary fixed value of $X$, the equilibrium density $\rho_{0,X}$ is derived by the condition
\begin{equation}
\left.\frac{\partial}{\partial\rho_{0}}E_{\mathrm{tot}}(\rho_{0},X)/A\right\vert_{A,\rho_{0}=\rho_{0,X}}=0.  \label{eqq1}
\end{equation}
Using Eqs. (\ref{e3}) and (\ref{eqq1}), we obtain the expression for the shift of bulk density (polarization 
effect) in the neutron rich nuclei 
\begin{equation}
\rho_{0,X}=\rho_{0,\mathrm{eq}}-9\frac{P_{A}}{K_{A}}(X-X^{\ast})^{2}.  
\label{rho1}
\end{equation}
The equilibrium partial pressure $P_A$ is positive and thereby $\rho_{0,X}<\rho_{0,\mathrm{eq}}$, see also Refs.
\cite{oyta98,oyii03}. We point out that in general the sign of the equilibrium partial pressure $P_A$ depends on 
the Skyrme force parametrization and this fact can be used for the Skyrme force selection \cite{brow01}. 

\section{Radii of nucleon distributions and neutron skin}

As noted above, the bulk density $\rho_{0,X}$ is smaller for neutron-rich nuclei, more neutrons should be pushed 
off to enrich the skin providing the polarization effect. Thus, the rms radius $\sqrt{\left\langle r_{n}^{2}\right\rangle}$ of neutron distribution does not necessarily obey the saturation condition $\sqrt{\left\langle r_{n}^{2}\right\rangle}\nsim A^{1/3}$. As a consequence, the nuclei with significant excess 
of neutrons exhibit neutron skin, i.e., they are characterized by larger radii for the neutron than for proton 
distributions. 
\begin{figure}
\begin{center}
\includegraphics*[scale=0.5,clip]{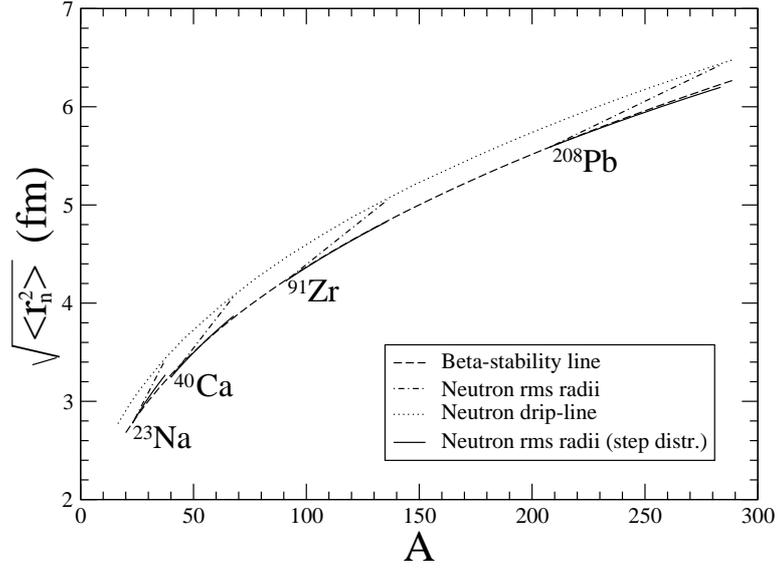}
\end{center}
\caption{The rms neutron radii beyond the beta-stability line for spherically symmetric nuclei $^{23}$Na, $^{40}$Ca,
$^{91}$Zr and $^{208}$Pb.}
\label{fig2}
\end{figure}
The neutron coat $N_S$ (see also \figurename\ \ref{fig1}) indicates the possibility of giant neutron halo which 
is growing with moving away from the beta stability line \cite{meto02}. In \figurename\ \ref{fig2} we have plotted 
the rms radii of neutron distribution from Eq. (\ref{rms}) as a function of $A$. The deviation of $\sqrt{\left\langle r_{n}^{2}\right\rangle}$ from the saturation behavior $\sim A^{1/3}$, obtained for the spherically symmetric nuclei,
demonstrates the appearance of giant neutron halo when approaching the drip line. To extract a simple geometrical
change of the radii we have made calculations with a step neutron distribution $\rho_n(r)=\rho_{0,n}\Theta(r-R_n)$,
where the radius of the neutron distribution has saturation behavior $R_n=r_{0,n}A^{1/3}$. The results of the
calculations are shown in \figurename\ \ref{fig2} by the solid lines. As one can see from \figurename\ \ref{fig2} 
the solid lines are very close to the beta-stability line. The difference between the dash-dotted and solid lines gives the value of the polarization effect. 

\begin{figure}
\begin{center}
\includegraphics*[scale=0.5,clip]{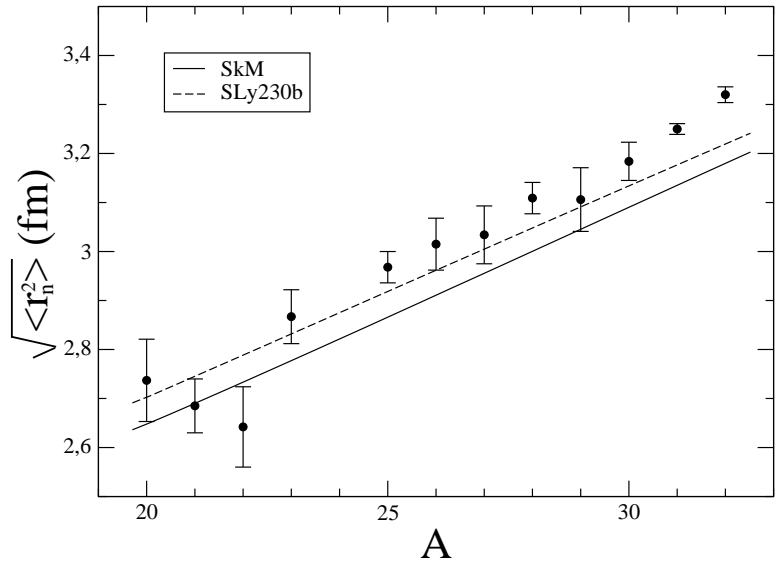}
\end{center}
\caption{The rms radius of neutron distribution in Na isotopes.}
\label{fig3}
\end{figure}

An occurrence of the giant halo in Na isotopes is shown in \figurename\ \ref{fig3}. We can see from this figure 
that the ETFA results agree quite well with the experimental data from \cite{suge95}. The sensitivity of $\sqrt{\left\langle r_{n}^{2}\right\rangle}$ calculation to the choice of the Skyrme force can be also seen. 

\begin{figure}
\begin{center}
\includegraphics*[scale=0.5,clip]{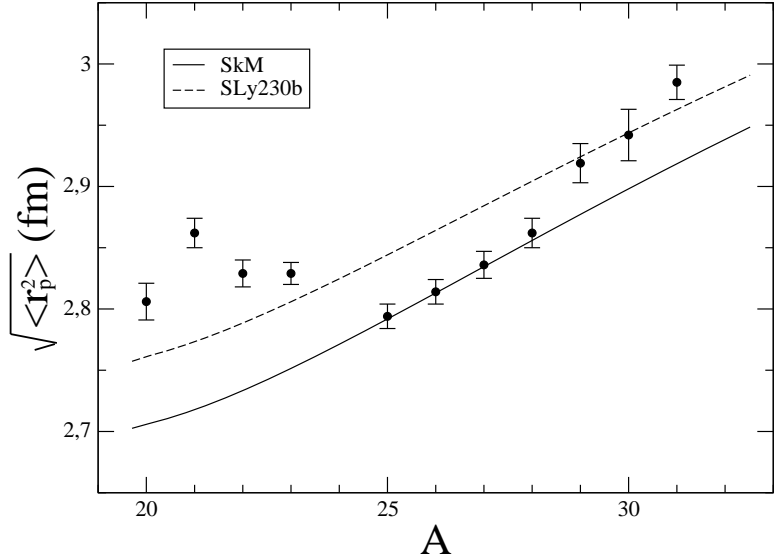}
\end{center}
\caption{The rms radius of proton distribution in Na isotopes.}
\label{fig4}
\end{figure}

The results for the charge radius $\sqrt{\left\langle r_{p}^{2}\right\rangle}$ are shown in \figurename\ \ref{fig4}. 
The rms radius $\sqrt{\left\langle r_{p}^{2}\right\rangle}$ of proton distribution indicates the non-monotonic
behavior due to the shell effects. Such behavior of $\sqrt{\left\langle r_{p}^{2}\right\rangle}$ correlates with 
$A$-dependence of the Coulomb radius $R_C$. To derive $R_C$ we use $E_{tot}(\rho_0,X)$ on the beta-stability line 
and establish an important relation for the chemical potentials $\lambda_{q}$ ($q=n$ for neutron and $q=p$ for 
proton) beyond the beta-stability line. Namely, for fixed $A$, we obtain
\begin{equation}
\Delta \lambda (X)
=\lambda_{n}-\lambda_{p}
=\left. \frac{\partial E}{\partial N}\right\vert_{Z}-\left. \frac{\partial E}{\partial Z}\right\vert_{N}
=2\left.\frac{\partial (E/A)}{\partial X}\right\vert_A
=4\left[ b_{\mathrm{sym}}^{\ast }(A)+e_{C}^{\ast }(A)\right] (X-X^{\ast}),  
\label{lambdaX}
\end{equation}
where $b_{\mathrm{sym}}^{\ast}(A)=b_{\mathrm{V,sym}}^{\ast}(\rho_{0,\mathrm{eq}})+b_{\mathrm{S,sym}}^{\ast}
(\rho_{0,\mathrm{eq}})\ A^{-1/3}$,
\begin{equation}
e_{C}^{\ast}(A)=0.15Ae^{2}/R_{C}
\label{rc}
\end{equation}
and
\begin{equation}
\lambda_{n}=\left(\frac{\partial E}{\partial N}\right)_{Z}, \quad
\lambda_{p}=\left(\frac{\partial E}{\partial Z}\right)_{N}.
\label{dlamb}
\end{equation}

On the beta-stability line, one has from Eq. (\ref{lambdaX}) that $\Delta\lambda (X)_{X=X^{\ast}}=0$, as it has 
to be from the definition of the beta-stability line. We point out that for finite nuclei, the condition 
$\Delta\lambda=0$ on the beta-stability line is not necessarily fulfilled exactly because of the discrete spectrum 
of single particle levels for both the neutrons and the protons near Fermi surface. In agreement with Eq.
(\ref{lambdaX}), the slopes of straight lines $\Delta\lambda(X)$ allow us to derive the quantity
$b_{\mathrm{sym}}^{\ast}(A)+e_{C}^{\ast}(A)$. From the beta-stability condition $\Delta\lambda(X)=0$ one can also
derive the asymmetry parameter $X^{\ast}(A)$. Using the definition of the beta-stability line as
$$
\left.\frac{\partial E_{\mathrm{tot}}(\rho_{0},X)/A}{\partial X}\right\vert_{A,\ X=X^{\ast}}=0,
$$
we obtain the symmetry energy coefficient $b_{\mathrm{sym}}^{\ast}(A)$ and Coulomb energy parameter
$e_{C}^{\ast}(A)$. Finally, using Eq. (\ref{rc}), we obtain the $A$-dependence of the Coulomb radius $R_{C}(A)$. 

The quantity $\partial (E/A)/\partial X$ in Eq. (\ref{lambdaX}) can be evaluated within the accuracy $\sim 1/A^{2}$
using the finite differences which are based on the experimental values of the binding energy per nucleon 
$\mathcal{B}(N,Z)=-E(N,Z)/A$. Namely, 
\begin{equation}
\left.\frac{\partial (E/A)}{\partial X}\right\vert_{A}
=\frac{A}{4}\,\left[\mathcal{B}(N-1,Z+1)-\mathcal{B}(N+1,Z-1)\right].  
\label{diff}
\end{equation}
Since the difference (\ref{diff}) is taken for $\Delta Z=-\Delta N=2$, the pairing effects do not affect the 
resulting accuracy. Because of Eq. (\ref{rc}), this procedure allows us to derive the "experimental" value of 
the Coulomb radius $R_{C}(A)$.
\begin{figure}
\begin{center}
\includegraphics*[scale=0.5,clip]{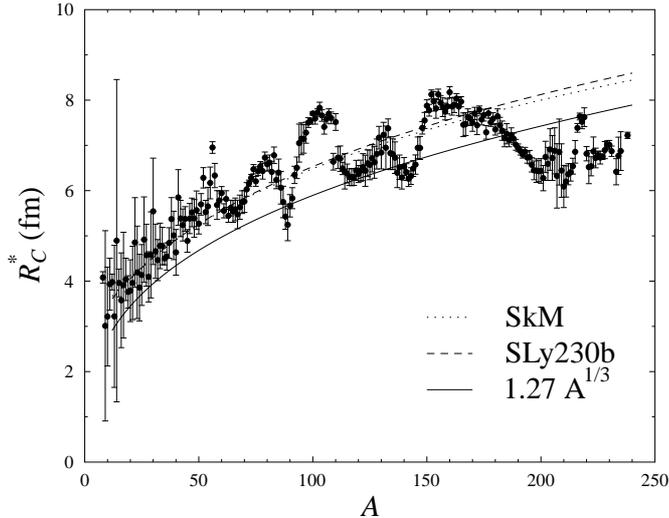}
\end{center}
\caption{The $A$-dependence of "experimental" Coulomb radius extracted from the experimental data \cite{awt03}.}
\label{fig5}
\end{figure}
The $A$-dependence of the "experimental" Coulomb radius $R_{C}^{\ast}(A)$, extracted from the experimental 
data \cite{awt03}, along the beta-stability line is presented in \figurename\ \ref{fig5}. The deviation of 
the Coulomb radius $R_{C}(A)$ from the smooth $A$-dependence is mainly due to the shell oscillations. 
We point out that the shell oscillations in $R_{C}^{\ast}(A)$ are correlated with the ones in $A$-dependence 
of the symmetry energy \cite{kosa09}.

\begin{figure}
\begin{center}
\includegraphics*[scale=0.5,clip]{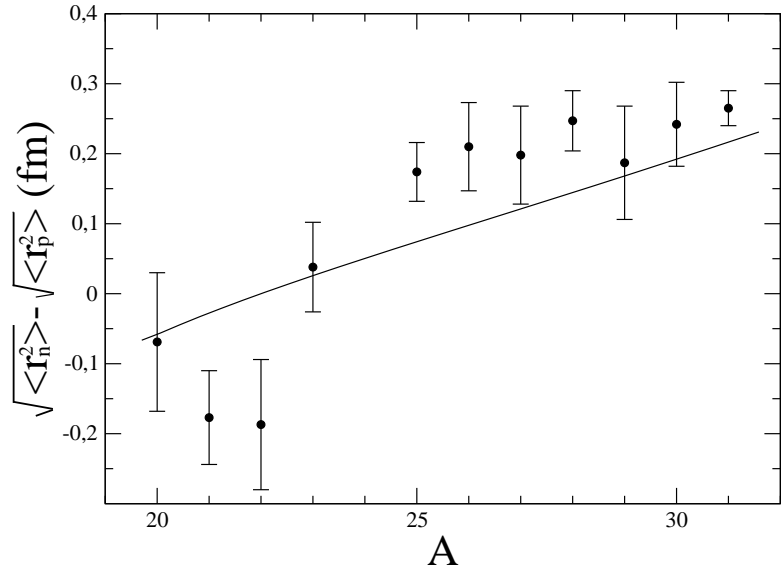}
\end{center}
\caption{Isovector shift of nuclear radius for Na isotopes.}
\label{fig6}
\end{figure}

The size of the neutron skin $\sqrt{\left\langle r_{n}^{2}\right\rangle}-\sqrt{\left\langle r_{p}^{2}\right\rangle}$ is illustrated in \figurename\ \ref{fig6}. The line has been obtained from Eq. (\ref{rms1}), and the experimental 
data were taken from Ref. \cite{suge95}. As seen from \figurename\ \ref{fig6}, the skin size 
$\sqrt{\left\langle r_{n}^{2}\right\rangle}-\sqrt{\left\langle r_{p}^{2}\right\rangle}$ is primarily linear with 
the asymmetry parameter $X$. 

\section{Summary}
We have applied the direct variational method within the extended Thomas-Fermi approximation with effective 
Skyrme-like forces to the description of the radii of nucleon distributions. In our consideration, the thin-skinned
nucleon densities $\rho_{p}(\mathbf{r})$ and $\rho_{n}(\mathbf{r})$ are generated by the profile functions which 
are eliminated by the requirement for the energy of the nucleus to be stationary with respect to variations of these
profiles. The advantage of the direct variational method is the possibility to derive the equation of state for 
finite nuclei: dependence of the binding energy per particle or the pressure on the bulk density $\rho_{0}$. 
We have evaluated the partial pressure $P_{A}$ which includes the contributions from the symmetry and the Coulomb
energies. The pressure $P_{A}$ is positive driving off the neutrons in neutron-rich nuclei to the skin. 

Using the leptodermous properties of the nucleon densities $\rho_{p}(\mathbf{r})$ and $\rho_{n}(\mathbf{r})$, 
we have established the possibility of giant neutron halo in neutron-rich nuclei. The effect of giant halo 
increases with moving away from the beta stability line. In \figurename\ \ref{fig2} this fact is demonstrated 
as deviation of the rms radii of neutron distribution from the saturation behavior $\sim A^{1/3}$ in the nuclei 
beyond the beta-stability line. 

The average behavior of the rms radii of nucleon distributions $\sqrt{\left\langle r_{q}^{2}\right\rangle}$ and 
the size of neutron skin are satisfactorily described within the extended Thomas-Fermi approximation. The sensitivity
of the calculations of  $\sqrt{\left\langle r_{q}^{2}\right\rangle}$ to force parametrization can be used for Skyrme
force selection. The charge radius of proton distribution shows the strong shell oscillations with mass number. 
We have demonstrated the relation of the Coulomb radius to the  isospin shift of the neutron-proton chemical potentials $\Delta\lambda=\lambda_{n}-\lambda_{p}$ for nuclei beyond the beta-stability line at fixed value of 
mass number $A$.

\end{document}